\documentclass[letterpaper, 10 pt, conference]{ieeeconf}
\IEEEoverridecommandlockouts 
\usepackage{arydshln}
\usepackage[T1]{fontenc}
\usepackage{caption,color}
\usepackage{graphicx}
\usepackage{amsmath,amssymb,amsfonts}
\usepackage{textcomp}
\usepackage{dsfont}
\usepackage{bbold}
\usepackage{arydshln}
\usepackage{comment}
\usepackage{mathtools} 
\usepackage{mathrsfs}
\usepackage{enumerate}

\everymath{\displaystyle}
\usepackage{cases}
\usepackage[framed,thmmarks]{ntheorem}  
\usepackage[overload]{empheq}
\usepackage{booktabs}
\usepackage{datetime}
\usepackage{cite}
\usepackage{stfloats}
\usepackage{pifont}
\newenvironment{proofa}{{\noindent\bfseries{Proof:}}}{\hfill$\blacksquare$}
\usepackage{etoolbox}

\newtheorem{proposition}{Proposition}
\newtheorem{lemma}{Lemma}
\newtheorem{theorem}{Theorem}
\newtheorem{assumption}{Assumption}
\newtheorem{definition}{Definition}
\AtEndEnvironment{problem}{\hfill$\Box$}
\AtEndEnvironment{lemma}{\hfill$\Box$}
\AtEndEnvironment{theorem}{\hfill$\Box$}
\AtEndEnvironment{assumption}{\hfill$\Box$}
\AtEndEnvironment{condition}{\hfill$\Box$}
\AtEndEnvironment{remark}{\hfill$\Box$}
\AtEndEnvironment{corollary}{\hfill$\Box$}
\AtEndEnvironment{stassumption}{\hfill$\Box$}
\AtEndEnvironment{definition}{\hfill$\Box$}

\title{\textbf{A Geometric Approach For Pose and Velocity Estimation Using IMU and Inertial/Body-Frame Measurements}}

\def\namedlabel#1#2{\begingroup
    #2%
    \def\@currentlabel{#2}%
    \phantomsection\label{#1}\endgroup
}
\author{
Sifeddine Benahmed, Soulaimane Berkane, \IEEEmembership{Senior Member, IEEE}, Tarek Hamel, \IEEEmembership{Fellow Member, IEEE}
\thanks{Sifeddine Benahmed is with the Department of Technology \& Innovation, Capgemini Engineering, Toulouse, 31300, France (sif-eddine.benahmed@capgemini.com).}
\thanks{Soulaimane Berkane is with the Department of Computer Science and Engineering, University of Quebec in Outaouais, QC J8X 3X7, Canada (soulaimane.berkane@uqo.ca).}
\thanks{Tarek Hamel is with I3S-UniCA-CNRS, University Cote d’Azur, and the Institut Universitaire de France, France
(thamel@i3s.unice.fr)}
\thanks{* This research work is supported in part by NSERC-DG RGPIN-2020-04759 and the Fonds de recherche du Qu\'ebec (FRQ).}
\thanks{}}

\theoremstyle{dotlessP}

\newcommand{\RNum}[1]{\uppercase\expandafter{\romannumeral #1\relax}}

\newcommand{\R}{\mathds{R}}

\newcommand{\Zsg}{\mathds{Z}_{>0}}

\newcommand{\sphere}{\mathds{S}}

\newcommand{\iframe}{\{\mathcal{I}\}}
\newcommand{\bframe}{\{\mathcal{B}\}}
\newcommand{\so}{\mathrm{SO}(3)}
\newcommand{\wx}{[\omega]_{\times}}
\newcommand{\Rthree}{\R^{3}}
\newcommand{\liealg}{\mathbf{so}(3)}
\newcommand{\posif}{p^{\mathcal{I}}}
\newcommand{\dposif}{\dot{p}^{\mathcal{I}}}
\newcommand{\velif}{v^{\mathcal{I}}}
\newcommand{\dvelif}{\dot{v}^{\mathcal{I}}}

\newcommand{\gravifr}{g^{\mathcal{I}}}

\newcommand{\accelbfr}{a^{\mathcal{B}}}

\newcommand{\id}{\mathbf{I}}

\newcommand{\vect}{\mathrm{vec}}
\newcommand{\by}{\mathbf{y}}

\newcommand{\setwo}{\mathrm{SE}_2(3)}
\newcommand{\sefive}{\mathrm{SE}_5(3)}
\newcommand{\am}{\alpha^{m}}

\newcommand{\av}{\alpha^{v}}
\newcommand{\mX}{\mathcal{X}}
\newcommand{\mhX}{\hat{\mathcal{X}}}
\newcommand{\mU}{\mathcal{U}}
\newcommand{\mG}{\mathcal{G}}
\newcommand{\mD}{\mathcal{D}}
\newcommand{\mE}{\mathcal{E}}

\newcommand{\tv}{\tilde{v}}
\newcommand{\tR}{\tilde{R}}

\newcommand{\tzB}{\tilde{z}^{\mathcal{B}}}
\newcommand{\tz}{\tilde{z}}

\begin{document}
\maketitle 
\begin{abstract}
This paper addresses accurate pose estimation (position, velocity, and orientation) for a rigid body using a combination of generic inertial-frame and/or body-frame measurements along with an Inertial Measurement Unit (IMU). By embedding the original state space, $\so \times \R^3 \times \R^3$, within the higher-dimensional Lie group $\sefive$, we reformulate the vehicle dynamics and outputs within a structured, geometric framework. In particular, this embedding enables a decoupling of the resulting geometric error dynamics: the translational error dynamics follow a structure similar to the error dynamics of a continuous-time Kalman filter, which allows for a time-varying gain design using the Riccati equation. Under the condition of uniform observability, we establish that the proposed observer design on $\sefive$ guarantees almost global asymptotic stability. We validate the approach in simulations for two practical scenarios: stereo-aided inertial navigation systems (INS) and GPS-aided INS. The proposed method significantly simplifies the design of nonlinear geometric observers for INS, providing a generalized and robust approach to state estimation.
\end{abstract}
\section{Introduction}
Inertial Navigation Systems (INS) are algorithms designed to estimate a vehicle’s navigation states, including position, velocity and attitude, relative to a fixed reference frame. The core sensor in INS is the Inertial Measurement Unit (IMU), which consists of a gyroscope and a 3-axis accelerometer that measure the vehicle’s angular velocity and specific acceleration, respectively. In an ideal scenario, where measurements are perfectly accurate and initial states are precisely known, the vehicle’s dynamics could be forward-integrated to determine its navigation states at any given time. However, in practice, sensor noise introduces errors that cause the state estimates to drift from the true values over time \cite{Woodman_INS_tech_report}. To mitigate this issue, modern INS algorithms incorporate additional measurements, such as those from the Global Positioning System (GPS), which provides periodic corrections to the position estimates. In GPS-denied environments, such as indoor scenarios, alternative sensors like vision or acoustic systems are employed to obtain additional measurements. For instance, vision-aided INS combines IMU data with visual inputs from cameras to enhance state estimation accuracy
\cite{Marco_ECC_2020, reis2018source, Single_bearing_ECC2024, wang2020hybrid}. This multi-sensor approach improves the overall reliability of the system by compensating for the limitations of individual sensors, ensuring robust and accurate navigation.

Traditional Kalman-type filters such as the Extended Kalman Filter (EKF) \cite{EKF_example} are effective for integrating multiple sensor inputs but are constrained by their reliance on local linearization, which makes them sensitive to initial estimation errors and less robust when dealing with nonlinear dynamics \cite{ReedJesse, multiple0trakingKim2022, EKF0Marina0application0Zhao}. Invariant EKF (IEKF) \cite{barrau2017invariant,barrau2018invariant} have emerged as a more robust and generic alternative, offering local asymptotic stability while addressing several limitations of traditional methods. By exploiting the geometric properties of Lie groups, the IEKF reduces sensitivity to initial conditions and preserves the system's natural geometric structure. Recent research has increasingly focused on developing nonlinear deterministic observers, which typically offer stronger stability guarantees and are more effective at addressing the intrinsic nonlinearities in INS applications, as seen, for instance, in \cite{mahony2008nonlinear,izadi2014rigid,zlotnik2016nonlinear, bryne2017nonlinear,Berkane_Automatica_2021, Wang_TAC_2022}.

Inertial navigation state variables—such as orientation \( R \), position \( p \), and velocity \( v \)—can be grouped into a single entity that belongs to the extended Special Euclidean group, $\setwo$, as shown in \cite{Barrai_Bonnabel_IEKF_stable}. While the invariant extended Kalman filter (IEKF) leverages this group structure, it provides only local asymptotic stability. Due to coupling in the estimation error dynamics (discussed in Section \ref{section:motivation}), achieving almost global asymptotic stability (AGAS) for general cases remains challenging. In specific cases, however, AGAS has been achieved, such as landmark measurements using a centroid-based homogeneous transformation to decouple rotational and translational error dynamics \cite{2020_TAC_hybrid_Wang}, and bearing measurements \cite{Wang_TAC_2022}, where auxiliary states were introduced.

In this paper, we propose a novel nonlinear geometric observer on \(\sefive\)  (instead of $\setwo$) that achieves AGAS for inertial navigation under generic measurements. First, inspired from \cite{Wang_TAC_2022}, we extend the original state with three auxiliary variables and demonstrate how inertial-frame measurements can be reformulated as body-frame relative measurements (see also \cite{ACC_2025_Universal_Observer}). This approach allows for the reformulation of the system dynamics and the generic measurements, fitting seamlessly within an observer design on the \(\sefive\) Lie group under right-invariant outputs. Interestingly, this embedding leads to a decoupling of the geometric error dynamics and enables the observer to integrate a wide range of inertial-frame and body-frame measurements, accommodating sensors such as GPS, landmarks, magnetometers, and much more. Additionally, we show that the "extended" translational error dynamics of the proposed observer follow similar trajectories to those of the linear Kalman filter's error dynamics, facilitating the gain design for the innovation term and sidestepping the complexities typically encountered in nonlinear observer design. We establish that the proposed observer guarantees AGAS. To the best of our knowledge, this is the first instance of an observer developed on \(\sefive\), offering a unified, adaptable framework that simplifies the design of nonlinear geometric observers for inertial navigation across various practical applications.

\section{Notation}\label{section:Notation_Preliminaries}
We denote by $\mathds{Z}_{>0}$  the set of positive integers, by $\R$ the set of reals, by $\R^n$ the $n$-dimensional
Euclidean space, and by $\sphere^n$ the unit $n$-sphere embedded in
$\R^{n+1}$. We use $\|x\|$ to denote the Euclidean
norm of a vector $x\in\R^n$, and $\textstyle \|X\|_F=\sqrt{\mathrm{trace}(X^{\top}X})$
to denote the Frobenius norm of a matrix $X\in\R^{n\times m}$. The $i$-th element of a vector $x\in\R^{n}$ is denoted by $x_i$. The $n$-by-$n$ identity and zeros matrices are denoted by $I_n$ and $0_{n\times n}$, respectively. The Special Orthogonal group of order three is denoted
by $\so:= \{A\in\R^{3\times3}: \mathrm{det}(A) = 1; AA^{\top} =A^{\top}A=I_3\}$. The set 
$\liealg:=\{\Omega\in\R^{3\times3}:\Omega=-\Omega^{\top}\}$ denotes the Lie algebra of $\so$. Let $n\in\Zsg$, the matrix Lie group $\mathrm{SE}_n(3)$ is defined as $\mathrm{SE}_n(3):=\{X=\mathcal{T}_n(R,x)\in \mathrm{SE}_n(3): R\in \so$ and $x\in\R^{3\times n}\}$, with the map $\mathcal{T}_n:\so \ltimes \R^{3\times n}\to \R^{(3+n)\times(3+n)}$ defined as
\begin{equation}
    \mathcal{T}_n(R,x)=\left[\begin{array}{c:c}
  \begin{matrix}
      R
  \end{matrix} &  x\\
  \hdashline \\  [-1em] 
  0_{n\times 3} & I_{n} \\
\end{array}
\right].
\end{equation}
The set $\mathbf{se}_n(3):=\{U=\mathcal{T}_n(\Omega,v):\Omega\in\liealg$ $v\in\R^{3\times n}\}$ denotes the Lie algebra associated to $\mathrm{SE}_n(3)$. The Lie bracket 
$[A,B]:=AB-BA$ for matrices 
$A$ and $B$ denotes the commutator of two matrices. The vector of all ones is denoted by $\mathds{1}$. For $x,\ y\in\Rthree$, the map $[.]_{\times}:\Rthree\to\liealg$ is defined such that $[x]_{\times}y=x\times y$ where $\times$ is the vector
 cross-product in $\Rthree$. Let $\mathrm{vex} : \liealg \to \R^3$ be the inverse isomorphism of the map $[\cdot]_{\times}$ such that
$\mathrm{vex}([\omega]_{\times}) = \omega$ for all $\omega\in\R^{3}$. The Kronecker product between two matrices $A$ and $B$ is denoted by $A
 \otimes B$. The  vectorization operator $\vect:\R^{m \times n}\to\R^{mn}$ , stacks the columns of a matrix $A \in \R^{m \times n}$ into a single column vector in $\R^{mn}$. The inverse of the vectorization operator, $\vect_{m,n}^{-1}:\R^{mn}\to\R^{m\times n}$, reconstructs the matrix from its vectorized form by reshaping the $mn\times1$ vector back into an $m\times n$ matrix form. For a matrix $A\in\R^{3\times 3}$, we denote by $\mathds{P}:\R^{3\times 3}\to\liealg$ the anti-symmetric projection of $A$ such that $\mathds{P}(A):=(A-A^{\top})/2$.  Define the composition map $\psi:=\mathrm{vex}\circ\mathds{P}$  such that, for a matrix $A=[a_{ij}]\in\R^{3\times3}$, one has $\textstyle\psi(A)=\frac{1}{2}[a_{32}-a_{23},a_{13}-a_{31},a_{21}-a_{12}]$.
 \section{Problem Formulation}\label{Section:Problem_formulation}
\subsection{Kinematic Model}
Let $\iframe$ be an inertial frame, $\bframe$ be an NED body-fixed frame attached to the center of mass of a rigid body (vehicle) and the rotation matrix $R\in\so$ be the orientation (attitude) of frame $\bframe$ with respect to $\iframe$. Consider the following 3D kinematics of a rigid body
\begin{subequations}\label{equation:dynamic_model_Inertial_frame}
\begin{align}
\label{eq:dp}
\dposif&=\velif,\\
\label{eq:dv}
\dvelif&=\gravifr+R\accelbfr,\\
\dot{R}&=R\wx,
\end{align}
\end{subequations}
where the vectors $\posif\in\R^{3}$ and $\velif\in\R^{3}$ denote the position and linear velocity of the rigid body expressed in frame $\iframe$, respectively, $\omega$ is the angular velocity of $\bframe$ with respect to $\iframe$ expressed in $\bframe$, $\gravifr\in\R^{3}$ is the gravity vector expressed in $\iframe$, and  $\accelbfr\in\R^{3}$ is the 'apparent acceleration' capturing all non-gravitational forces applied to the rigid body expressed in frame $\bframe$. 

This work focuses on the problem of position, linear velocity, and attitude estimation for Inertial Navigation Systems (INS). We assume that the vehicle is equipped with an Inertial Measurement Unit (IMU) providing measurements of the angular velocity $\omega$ and apparent acceleration $\accelbfr$ (inputs). Note that the translational system \eqref{eq:dp}-\eqref{eq:dv} is a linear system with an unknown input $R\accelbfr$. Therefore, there is a coupling between the translational dynamics and the rotational dynamics through the accelerometer measurements. Most adhoc methods in practice assume that $R\accelbfr\approx -\gravifr$ to remove this coupling between the translational and rotational dynamics. However, this assumption holds only for non-accelerated vehicles, {\it i.e., } when $\dvelif\approx 0$. In this work, we instead design our estimation algorithm without this latter assumption.

\subsection{Objective}
The objective of this paper is to design an almost global asymptotic convergent observer to simultaneously estimate the inertial position $\posif$, inertial velocity $\velif$ and attitude $R$ using \textit{all or part} of the following generic outputs:
\begin{assumption}[Generic Outputs]\label{assumption:available_measurements}
We assume that all or part of the following measurements are available:\\
$\quad (i)$ The body-frame measurements $\eta^{\mathcal{B}}_i=R^{\top}(\xi_i^{\mathcal{I}}-\gamma_ip^{\mathcal{I}})$ where  $\xi_i^{\mathcal{I}}\in\R^{3}$ constant and known and $\gamma_i\in\{0,1\}$, $i\in\{1,\cdots,p\}$, $p\in\Zsg$, \textbf{and/or}\\
$\quad (ii)$ The inertial-frame measurements $\boldsymbol{\eta}_i^{\mathcal{I
}}=\posif+Rb_i$. where $b_i\in\R^{3}$ constant and known, $i\in\{1,\cdots,q\}$, $q\in\Zsg$, \textbf{and/or}\\
$\quad (iii)$ The inertial-frame linear velocity $\eta_{v}^{\mathcal{I}}=\velif$, \textbf{and/or}\\
$\quad (iv)$ The body-frame linear velocity $\eta_{v}^{\mathcal{B}}=R^{\top}\velif$.
\end{assumption}
The measurements described in items $(i)-(iv)$ represent general outputs that can be derived from various sensor configurations depending on the application. Item $(i)$ with $\gamma_i = 1$ corresponds to body-frame landmark measurements (\textit{e.g.,} from a stereo vision system). When $\gamma_i = 0$, the measurements simplify to observations of a known and constant inertial vector in the body-frame (\textit{e.g.,} from a magnetometer). Item $(ii)$ captures for instance position measurements from a GPS receiver with lever arm $b_i$. Items $(iii)$ and $(iv)$ correspond to velocity measurements either in inertial-frame (\textit{e.g.,} from GPS) or in body-frame  (\textit{e.g.,} from airspeed sensor or Doppler radar). Finally, note that the measurement models in Assumption \ref{assumption:available_measurements} did not include bearing measurements (\textit{e.g.,} from monocular cameras) to simplify the exposition but this can be included easily using projection operators; see \cite{Wang_TAC_2022}.
\section{Main Result}\label{section:main_result}
In this section, we present the main result. First, we define a \textit{processed} output vector based on the generic measurements outlined in Assumption~\ref{assumption:available_measurements}, ensuring the resulting system is compatible with the $\sefive$ framework. Following this, we introduce the proposed observer design, detail the innovation term, and provide a convergence analysis.
\subsection{Motivation}\label{section:motivation}
As shown in \cite{barrau2017invariant}, the dynamic variables $\posif$, $\velif$ and $R$ can be grouped in a single element $\mX:=\mathcal{T}_2(R,\posif,\velif)$ that belongs to the Lie group $\setwo$. Following the work in \cite{Pieter2023_sychronous}, the kinematics \eqref{equation:dynamic_model_Inertial_frame} can be written in a compact form as:
\begin{equation}\label{equation:dynamics_in_SE_2}
   \dot{\mX}= \mX\mU+\mG\mX+[\mD,\mX],
\end{equation}
where 
\begin{align}
    \mU&:=\left[\begin{array}{c:c}
 \wx  & \begin{matrix}0_{3\times 1} &\accelbfr \end{matrix} \\
  \hdashline \\  [-1em] 
  0_{2\times 3}  & 0_{2\times 2} \\
\end{array}
\right], \mG:=\left[\begin{array}{c:c}
 0_{3\times 3}  & \begin{matrix}0_{3\times 1} &\gravifr\end{matrix} \\
  \hdashline \\  [-1em] 
  0_{2\times 3}  & 0_{2\times 2} \\
\end{array}
\right],\nonumber\\
\mD&:=\left[\begin{array}{c:c}
 0_{3\times 3}  & 0_{3\times 2} \\
  \hdashline \\  [-1em] 
  0_{2\times 3}  & \begin{matrix}0 & 0 \\ -1&0\end{matrix} \\
\end{array}
\right].\nonumber
\end{align}
Let $\hat{p}^{\mathcal{I}}$, $\hat{v}^{\mathcal{I}}$ and $\hat{R}$ be the estimates of $\posif$, $\velif$ and $R$, respectively, and $\mhX:=\mathcal{T}_2(\hat{R},\hat{p}^{\mathcal{I}},\hat{v}^{\mathcal{I}})$. Consider the following \textit{pre-observer}, which copies the dynamics of \eqref{equation:dynamics_in_SE_2} 
\begin{equation}\label{equation:dynamics_observer_in_SE_2}
   \dot{\mhX}= \mhX\mU+\mG\mhX+[\mD,\mhX].
\end{equation}
Now, define the right-invariant estimation error $\mE$ as
\begin{equation}
    \mE=\mX\mhX^{-1}.
\end{equation}
In view of \eqref{equation:dynamics_in_SE_2} and \eqref{equation:dynamics_observer_in_SE_2}, the geometric error dynamics are given by
\begin{equation}\label{equation:dynamic_of_error_in_SE_2}
    \dot{\mE}=\left[\begin{array}{c:c}
 0_{3\times 3}  & \begin{matrix}
     \tv&(I_3-\tR)\gravifr
 \end{matrix} \\
  \hdashline \\  [-1em] 
  0_{2\times 3}  & 0_{2\times 2} \\
\end{array}
\right],
\end{equation}
where $\tR:=R\hat{R}^{\top}$ and $\tv:=\velif-\tR\hat{v}^{\mathcal{I}}$   represent the geometric estimation errors of the attitude and the velocity, respectively. One can see clearly that the coupling between the attitude and translational geometric errors is evident 
because of the term $(I_3-\tR)\gravifr$. This coupling complicates the design of globally convergent observers, as the attitude error affects the translational error through gravity. Most existing estimation approaches, such as the Invariant Extended Kalman Filter (IEKF)\footnote{Note that the IEKF proposed in \cite{barrau2017invariant} can be used here because the system \eqref{equation:dynamic_model_Inertial_frame} satisfies the group affine property.} \cite{barrau2017invariant}, rely on local linearization and thus guarantee only local convergence.

To remove this coupling, we  extend the state of the system with the auxiliary state $w:=[e_1^{\top},e_2^{\top},e_3^{\top}]^{\top}$, which satisfies
\begin{equation}
    \dot{w}=0_{9\times1},
\end{equation}
leading to a new compact state matrix 
\begin{equation}
  X:=\mathcal{T}_5(R,p,v,e_1,e_2,e_3)\in\sefive,
\end{equation}which has the following dynamics
\begin{equation}\label{equation:dynamics_in_SE_5_3}
    \dot{X}=XU+[X,D],
\end{equation}
where the matrices $D$ and $U$ are written as follows,
\begin{flalign}\label{equation:D_U_Abar}
 \hspace{1cm} D&:=\left[\begin{array}{c:c}
 0_{3\times 3}  & 0_{3\times 5}  \\
  \hdashline \\  [-1em] 
  0_{5\times 3}  & \bar{A}^{\top} \\
\end{array}
\right],  U:= \left[\begin{array}{c:c}
\wx & \begin{matrix}
    0 & \accelbfr &0_{3\times 3} 
\end{matrix}\\
  \hdashline \\  [-1em] 
  0_{5\times 3} &0_{5\times 5} 
\end{array}
\right],\nonumber\\
\bar{A}
&=\begin{bmatrix}0 &1 &0_{1\times3} \\
0&0&(\gravifr)^{\top}\\
0_{3\times1} & 0_{3\times1} &0_{3\times3}
\end{bmatrix}\in\mathbb R^{5\times 5}.
\end{flalign}
Now, let $\hat{X}$ be the estimates of $X$ and consider the following {\it pre-observer}, which copies the dynamics of \eqref{equation:dynamics_in_SE_5_3},
\begin{equation}\label{equation:dynamics_observer_in_SE_5}
     \dot{\hat{X}}=\hat{X}U+[\hat{X},D],
\end{equation}
and define the following right-invariant geometric error
\begin{equation}\label{equatioin:right_invariant_geometric_error}
    E=X\hat{X}^{-1}.
\end{equation}
Let $\hat{w}$ be the estimate of $w$ and $\tilde{w}:=w-(I_3\otimes\tR)\hat{w}$, which represents the geometric estimation error of $w$. The dynamic of $E$ is given by:
\begin{equation}\label{equation:dynamic_of_error_in_SE_2}
    \dot{E}=\left[\begin{array}{c:c}
 0_{3\times 3}  & \begin{matrix}
     \tv&((\gravifr)^{\top}\otimes I_3)\tilde{w} & 0_{3\times 3}
 \end{matrix} \\
  \hdashline \\  [-1em] 
  0_{5\times 3}  & 0_{5\times 5} \\
\end{array}
\right].
\end{equation}
The introduction of the auxiliary state 
$w$ effectively decouples the geometric errors of the attitude and translational variables. While the idea of employing an auxiliary state $w$ has been explored in prior works, notably in \cite{Wang_TAC_2022}, it was limited to the context of bearing-to-landmark measurements. In contrast, this paper extends the approach to the Lie group $\sefive$, incorporating a broader range of generic measurements obtained from various sensor types. Note that the Lie group $\sefive$ represents a novel geometric framework that has not been previously explored in the literature, providing new insights into observer design for inertial navigation systems.
\subsection{Pre-Processing of Inertial- and Body-Frame Measurements: Unified Formulation}
Under Assumption \ref{assumption:available_measurements}, let us define the following measurement vector: 
\begin{flalign}\label{equation:vector_of_all_measurements}
 \hspace{-0.5cm}   y=\begin{bmatrix}
        y_1 \\
        \vdots\\
        y_p \\
       y_{p+1} \\
        \vdots\\
       y_{p+q} \\
       y_{p+q+1} \\
        y_{m} 
    \end{bmatrix}
:=\begin{bmatrix}
       \eta_1^{\mathcal{B}}\\
        \vdots\\
       \eta_p^{\mathcal{B}} \\
      R^{\top}(\boldsymbol{\eta}_1^{\mathcal{I}}-\posif) \\
        \vdots\\
     R^{\top}(\boldsymbol{\eta}_q^{\mathcal{I}}-\posif)\\
        R^{\top}(\eta_v^{\mathcal{I}}-\velif) \\
       \eta_v^{\mathcal{B}}\\
    \end{bmatrix}=\begin{bmatrix}
         \eta_1^{\mathcal{B}}\\
        \vdots\\
       \eta_p^{\mathcal{B}} \\
      b_1 \\
        \vdots\\
        b_q \\
        0_{3\times 1} \\
       \eta_v^{\mathcal{B}} \\
    \end{bmatrix}.
\end{flalign}
Note that, if a measurement model is not available, the corresponding line in $y$ will be omitted. The next proposition rewrites the measurements in \eqref{equation:vector_of_all_measurements}  into a structure appropriate for $\sefive$ framework.
\begin{proposition}\label{proposition:reformuulation_of_the_measurements}
For any $i\in\mathcal{M}:=\{1,\cdots,m\}$, where $m=\mathbf{card}(y)/3$, there exists a reference vector (possibly time-varying) $r_i\in\R^5$, such that the measurements in \eqref{equation:vector_of_all_measurements} can be reformulated as follows:
    \begin{equation}\label{equation:universal_form_for_output_in_SEfive}
    \by_i=X^{-1}\mathbf{r_i},
    \end{equation}
    where $\mathbf{y}_i=\begin{bmatrix}
        y_i^{\top} & r_i^{\top}
    \end{bmatrix}^{\top}$  and $\mathbf{r}_i=\begin{bmatrix}
        0_{3\times1} & r_i^{\top}
    \end{bmatrix}^{\top}$.
\end{proposition}
\begin{proofa} See Appendix~\ref{appendix:proof_of_proposition:reformuulation_of_the_measurements}.

\end{proofa}

Proposition~\ref{proposition:reformuulation_of_the_measurements} shows that by expressing the system within the Lie group  $\sefive$ and thanks to the auxiliary states $e_1$, $e_2$ and $e_3$, both inertial- and body-frame outputs can be written in a unified formulation suitable to this framework. This novel approach enables the inclusion of a broader set of measurements, thereby enhancing the applicability of the proposed observer design. The vectors $r_i$, $i\in\mathcal{M}$, for each output are given in Table~\ref{Table:vector_r_for_each_measurement}.
\begin{table}[]
    \centering
    \begin{tabular}{c|c|c}
     Generic Measurements &   $y_i$ & $r_i$ \\
        \midrule
   item $(i)$ & $\eta_i^{\mathcal{B}}$     & $\begin{bmatrix}
          \gamma_i &0 &-(\xi_i^{\mathcal{I}})^{\top}
      \end{bmatrix}^{\top}$ \\
   item $(ii)$ &  $b_{i}$ 
     & $\begin{bmatrix}
           1 &0&-(\boldsymbol{\eta}_i^\mathcal{I})^{\top}
      \end{bmatrix}^{\top}$\\
    item $(iii)$ &  $0_{3\times1}$  & $\begin{bmatrix}
          0 &1&-(\eta_v^{\mathcal{I}})^{\top}
      \end{bmatrix}^{\top}$\\
    item $(iv)$ &  $\eta_v^{\mathcal{B}}$& $\begin{bmatrix}
           0 &-1&0_{1\times3}
      \end{bmatrix}^{\top}$
    \end{tabular}
    \caption{The reduced reference vector associated to each measurement in Assumption \ref{assumption:available_measurements}.} 
\label{Table:vector_r_for_each_measurement}
\end{table}
\subsection{Nonlinear Observer Design  on $\sefive$}
Let $z:=\begin{bmatrix}
p^{\mathcal{I}}&v^{\mathcal{I}}& e_1 & e_2 &e_3    
  \end{bmatrix}\in\R^{3\times5}$, $\hat{z}$ its estimates and $\tilde{z}=z-\tilde{R}\hat{z}$. The proposed innovation term associated to each measurement is as follows:
\begin{align}
    \Delta \mathbf{y}_i&:=\mathbf{r}_i(t)-\hat{X}X^{-1}\mathbf{r}_i(t)\\
    &=(I_8-\tilde{X}^{-1})\mathbf{r}_i(t),
\end{align}
and define $\Delta_z=\begin{bmatrix}
    (G\Delta \mathbf{y}_1)^{\top} & \cdots &  (G\Delta \mathbf{y}_m)^{\top}
\end{bmatrix}^{\top}$ where $G=\begin{bmatrix}
    I_3 & 0_{3\times5}
\end{bmatrix}$. 

We propose the following observer with a copy of the dynamics in \eqref{equation:dynamics_in_SE_5_3} and an innovation term on the matrix Lie group $\sefive$:
\begin{equation}\label{equation:observer_SE2_5_with_delta}
    \dot{\hat{X}}=\hat XU+[\hat X,D]+\Delta\hat X,
\end{equation}
where the innovation term $\Delta$  is given by,
\begin{align}\label{equation:DELTA}
\Delta:=\left[\begin{array}{c:c}
 [\Delta_R]_\times  & 
   \begin{matrix}
      K_p\Delta_z\; K_v\Delta_z \;K_{e1}\Delta_z  \;K_{e2}\Delta_z\;K_{e3}\Delta_z
   \end{matrix} \\
  \hdashline \\  [-1em] 
  0_{5\times 3}  & 0_{5\times 5} \\
\end{array}
\right],
\end{align}
where $ [\Delta_R]_{\times} \in\liealg$ and $K_p ,K_v,K_{e1},K_{e2},K_{e3}\in\R^{3\times 3m}$ are time-varying gains. The structure of the proposed approach is given in Fig.~\ref{figure:Illustration_of_the_proposed_method}.

 Let $\hat{e}_i$ be the estimates of $e_i$ for any $i\in\{1,2,3\}$, respectively. A possible choice for $\Delta_R$ is inspired from \cite{wang2021nonlinear} as follows:
\begin{equation}
    \Delta_R=\frac{1}{2}\sum_{i=1}^3\rho_i\hat e_i\times e_i,\quad \rho_i>0,
\end{equation}
where $\rho_i$, $i\in\{1,2,3\}$ are constant positive scalars.  Note that, typically, attitude estimation can be achieved using body-frame measurements of at least two non-collinear inertial-frame vectors \cite{mahony2008nonlinear}. However, when dealing with generic outputs, this minimal requirement may not always be satisfied with the available set of measurements, making it challenging to construct an appropriate innovation term $\Delta_R$. To address this challenge, the inertial basis vectors $e_i$   and their corresponding body-frame vectors  $R^{\top}e_i$   are utilized, as proposed in \cite{2020_TAC_hybrid_Wang}. Since 
$R^{\top}e_i$   is unknown, adaptive auxiliary vectors 
$\hat{e}_i$   are introduced, designed so that 
$\hat{R}^{\top}\hat{e}_i$   converges asymptotically to 
$R^{\top}e_i$.

In view of \eqref{equation:dynamics_in_SE_5_3}, \eqref{equatioin:right_invariant_geometric_error}, \eqref{equation:observer_SE2_5_with_delta} and  \eqref{equation:DELTA}, we obtain the following  geometric autonomous error dynamics
\begin{align}\label{equation:dynamics_geometric_error_with_innovation_term}
 \dot{E}&=[E,D]- E\Delta \nonumber\\&=\left[\begin{array}{c:c}
-\tilde R[\Delta_R]_\times  & \tilde z\bar{A}^{\top}-\tilde R K (I_5\otimes \Delta_z) \\
  \hdashline \\  [-1em] 
  0_{5\times 3}  & 0_{5\times 5}  \\
\end{array}
\right],
\end{align}
where  $K=\begin{bmatrix}
       K_p &K_v&K_{e1}&K_{e2}&K_{e3}
   \end{bmatrix}$ and $\bar{A}$ is given in \eqref{equation:D_U_Abar}. Note that the error
dynamics are independent of both the system’s trajectory
and the input, a desirable property found in geometric
observers on Lie groups, see for instance \cite{Barrai_Bonnabel_IEKF_stable,Lageman_Mahonny_Gradient_Like}. Next, we show that the dynamics of $E$ are decoupled  which is a feature that will allow us to use Riccati gain update for the linear subsystem $ \tilde z\bar{A}^{\top}-\tilde R K (I_5\otimes \Delta_z)$. 
\begin{figure*}[h!]
    \centering    \includegraphics[width=0.8\textwidth]{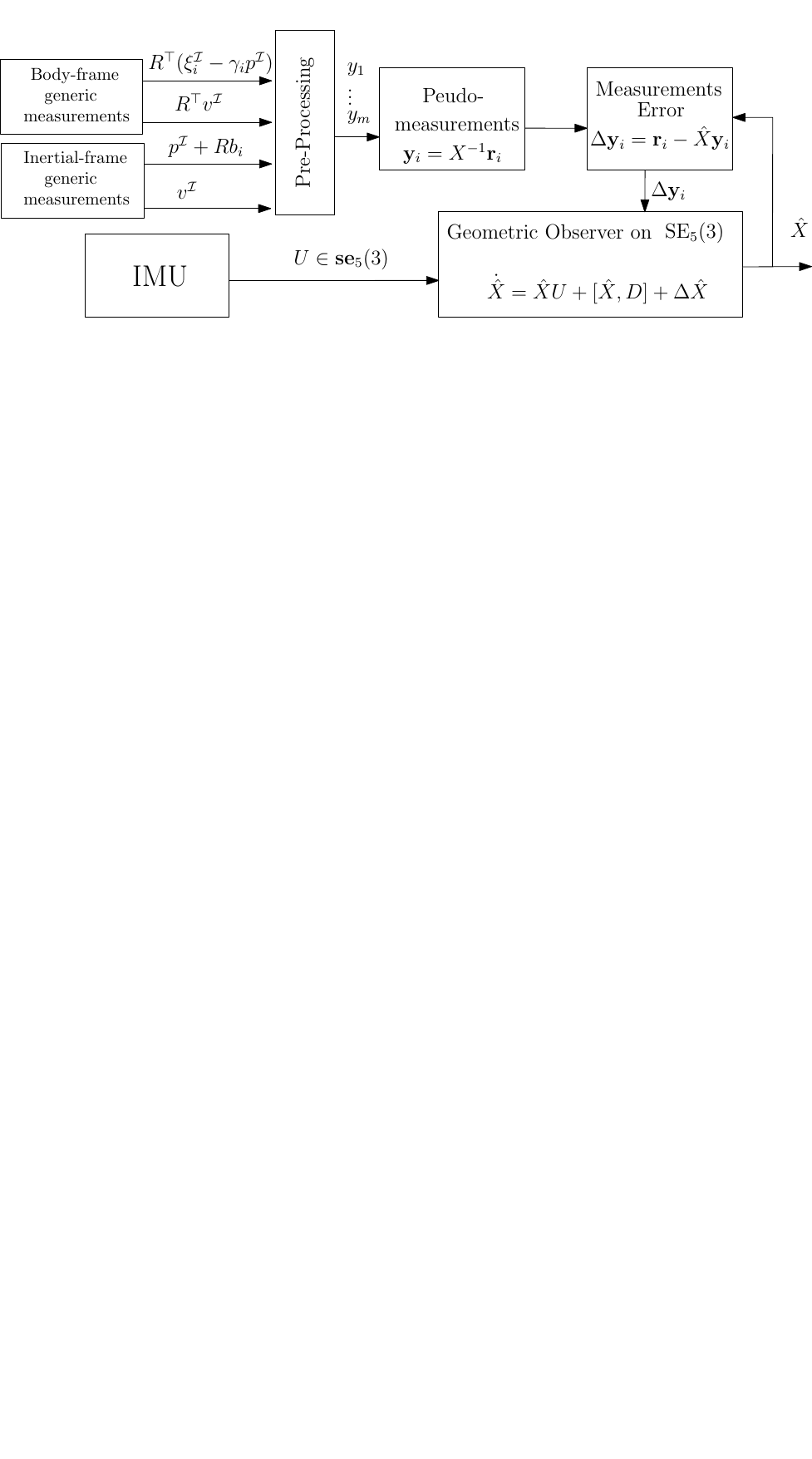}
    \caption{Illustration of the proposed geometric estimation approach}
    \label{figure:Illustration_of_the_proposed_method}
\end{figure*}
   \subsection{ Translational Error Dynamics  and Gain Design}
 To simplify the design of the time-varying gain $K(t)$, we start by rewriting the dynamics of the translational estimation error as stated in the following lemma.
\begin{lemma}\label{lemma:dynamics_of_the_translaional_error_in_the_body_frame}
    Let  $\tzB:=R^{\top}\tz$, $\tilde{x}^{\mathcal{B}}:=\vect(\tzB)$ and $K^{\mathcal{I}} =\begin{bmatrix}
       K_p^{\top}&K_v^{\top}&K_{e1}^{\top}&K_{e2}^{\top}&K_{e3}^{\top}
   \end{bmatrix}^{\top}$, the dynamics of $\tilde{x}^{\mathcal{B}}$ is given by
\begin{equation}\label{equation:closed_loop_dynamic_of_x_B}     \dot{\tilde{x}}^{\mathcal{B}}=(A(t)+K^{\mathcal{B}}(t)C(t))\tilde{x}^{\mathcal{B}},
   \end{equation}
   where $K^{\mathcal{B}}(t)=(I_5\otimes \hat{R}^{\top})K^{\mathcal{I}}(t)(I_m\otimes\hat{R})$ and
   $C(t)=\begin{bmatrix}
 (r_1^{\top}(t)\otimes I_3)^{\top}&\cdots&
 (r_m^{\top}(t)\otimes I_3)^{\top}
\end{bmatrix}^{\top}$ and where $A(t)=\bar{A}\otimes I_3+S(t)$ and $S(t)=
-I_5\otimes\wx$.
\end{lemma}
\begin{proofa}
See Appendix~\ref{appendix:proof_of_lemma_dynamics_of_the_translaional_error_in_the_body_frame}  
   \end{proofa}
   
   It is important to note that the matrix  $A(t)$ in Lemma~\ref{lemma:dynamics_of_the_translaional_error_in_the_body_frame} is time-varying because it depends on the profile of the angular velocity $\omega(t)$, which acts as an external time-varying signal. Similarly, the matrix $C(t)$ is time-varying, as it relies on the measurements of the inertial-frame position and velocity, both of which change over time.    
   Therefore, we impose the following realistic constraint on the translational estimation error's trajectory which is needed to ensure that the matrices $A(t)$ and $C(t)$ are well-conditioned for the convergence guarantees of the proposed observer.
\begin{assumption}\label{assumption::bounded_bw}
The time-varying matrices $A(t)\in\R^{n\times n}$ and $C(t)\in\R^{p \times n}$ are assumed continuously differentiable and uniformly bounded with bounded derivatives.
\end{assumption}

  Furthermore, Lemma~\ref{lemma:dynamics_of_the_translaional_error_in_the_body_frame} demonstrates that the translational error dynamics are decoupled from the attitude estimation error, and its trajectory follows the trajectory of a continuous-time Kalman filter. Consequently, the gain 
$K^{\mathcal{B}}(t)$ can be determined as follows:
\begin{equation}\label{equation:gain_general_Riccati_observer}
  K^\mathcal{B}(t)=PC(t)^{\top}Q(t),  
\end{equation}
where $P$ is the solution of the following Riccati equation:
\begin{equation}\label{equation:Riccati_equation}
    \dot{P}=A(t)P+PA^{\top}(t)-PC^{\top}(t)Q(t)C(t)P+V(t),
\end{equation}
and where $P(0)$ is a positive definite matrix and  $Q(t)$ and $V(t)$ are uniformly positive definite matrices that should be specified. Once the gain $K^{\mathcal{B}}$ is computed, the corresponding gain $K(t)$ can be derived accordingly. Note that, in the context of Kalman filter, the matrices $V(t)$ and $Q^{-1}(t)$ represent covariance matrices characterizing additive noise on the system state. 
\subsection{Uniform Observability and Almost Global Asymptotic Stability}\label{subsection:convergence_analysis}
The following definition formulates the well-known uniform observability condition in terms of the observability Gramian matrix. The uniform observability property guarantees uniform global exponential stability of the translational error dynamics \eqref{equation:K_BC(t)x_B}, see \cite{Hamel2017PositionMeasurements} for more details.
\begin{definition}\textbf{(Uniform Observability)}\label{definition:uniform_observability} The pair $(A(t)$$,C(t))$ is uniformly  observable if there exist constants $\delta,\mu>0$ such that $\forall t\geq0$
\begin{equation}\label{equation:conditon_of_uniform_observability}
    W(t,t+\delta):=\frac{1}{\delta}\int_t^{t+\delta}\hspace{-0.5em}\phi^{\top}(s,t)C^{\top}(s)C(s)\phi(s,t)ds\geq\mu I_n,  
\end{equation}
where $\phi(s,t)$ is the transition matrix associated to $A(t)\in\R^{n\times n}$ such that $\frac{d}{dt}\phi(t,s)=A(t)\phi(t,s)$ and $\phi(t,t)=I_n$.
\end{definition}

Sufficient conditions for uniform observability of the pair $(A(t),C(t))$ have been established in the literature for various practical scenarios, as seen in \cite{ACC_2025_Universal_Observer,Single_bearing_ECC2024,Wang_TAC_2022}. In the case of GPS-aided INS, for example, when the available measurements are those corresponding to items $(i)$-$(iii)$ of Assumption~\ref{assumption:available_measurements}.  Specifically, the measurements in item $(i)$ correspond to magnetometer readings in the body frame, i.e.,  $\gamma_1=0$ and $R^{\top}\xi_1^{\mathcal{I}}$ represents the magnetic field in the body-frame, and we assume one measurement from item $(ii)$ corresponding to GPS position with lever arm $b_1$. Under this measurements configuration, a sufficient condition for uniform observability of the corresponding pair $(A(t),C(t))$ is given in the following lemma adapted from \cite[Lemma 4]{ACC_2025_Universal_Observer}.

\begin{lemma}\label{Lemma:sufficient_condition_for_UO_GPS_INS}
Let $\xi_1^{\mathcal{I}}$ represents the magnetic field in the inertial frame and $\am,\av\in\{0,1\}$. If there exist $\delta,\mu>0$ such that, for any $t\geq0$,
\begin{flalign}\label{equation:condition_of_lemma_GPS_INS}
  & \frac{1}{\delta}\int_t^{t+\delta}( \dot{v}^{\mathcal{I}}(\tau)-\gravifr)(\dot{v}^{\mathcal{I}}(\tau)-\gravifr)^{\top}d\tau\nonumber\\
    &+\am\xi_1^{\mathcal{I}}(\xi_1^{\mathcal{I}})^{\top} + \frac{\av}{\delta}\int_t^{t+\delta}\velif(\tau)(\velif(\tau))^{\top}d\tau\geq\mu I_3,
\end{flalign}

then, the pair $(A(\cdot),C(\cdot))$ is uniformly observable.
\end{lemma}
Note that when $\am=1$ and $\av=1$, condition \eqref{equation:condition_of_lemma_GPS_INS} is generally satisfied, for instance even if $\dot{v}^{\mathcal{I}}=0$, the condition is met if gravity vector $\gravifr$ and the magnetic field $\xi_1^{\mathcal{I}}$ are non collinear, and the velocity $\velif$ does not consistently lie in the plane spanned by $\gravifr$ and $\xi_1^{\mathcal{I}}$. Now, when ($\am=1$ and $\av=0$), condition \eqref{equation:condition_of_lemma_GPS_INS} represents a persistent of excitation on the apparent acceleration $\dot{v}^{\mathcal{I}}-\gravifr$. Moreover, when $\am=0$, $\av=1$ and $\dot{v}^{\mathcal{I}}=0$, condition \eqref{equation:condition_of_lemma_GPS_INS} represents a Persistent of Excitation (PE) on the velocity $v^{\mathcal{I}}$. Overall, the PE condition ensures that the vehicle's trajectory generates sufficiently rich data to estimate its full state (position, velocity, and orientation).

The next theorem exploits the structure of the geometric estimation error to establish sufficient conditions that guarantee AGAS for the proposed observer design.
\begin{theorem}\label{theorem_AGAS}
Suppose that the pair $(A(t),C(t))$ is uniformly observable, design $K^{\mathcal{B}}$ as in \eqref{equation:gain_general_Riccati_observer}, and pick three distinct scalars $\rho_i>0$, $i\in\{1,2,3\}$. Then, the desired equilibrium point $E^{\star}=\mathcal{T}_5(I_3,0_{15\times 1})$ of \eqref{equation:dynamics_geometric_error_with_innovation_term} is almost globally asymptotically stable.
\end{theorem}
\begin{proofa}
See Appendix~\ref{appendix:proof_of_theorem_AGAS}.
\end{proofa}

Theorem~\ref{theorem_AGAS} exploits the  globally exponentially convergent property of the translational errors dynamics \eqref{equation:closed_loop_dynamic_of_x_B} and the  ISS property of the reduced attitude estimation error  to show that the interconnection preserves AGAS of the estimation errors.  
\section{Simulation Results}\label{section:simulation}
In this section, we provide simulation results to test the performance of the observer proposed in Section~\ref{section:main_result}. We consider two practical applications: Stereo-aided-INS and GPS-aided-INS. For the Stereo-aided-INS scenario, the available measurements are limited to item (i) of Assumption~\ref{assumption:available_measurements}. Specifically, we assume that we have a family of $5$ landmark measurements $\eta_i^{\mathcal{B}}=R^{\top}(\xi_i-\gamma_i\posif)$ with $\gamma_i=1$, $\forall i\in\{1,2,3,4,5\}$. To guarantee the uniform observability of the corresponding pair $(A(t),C(t))$, the configuration of the landmarks is carefully chosen to satisfy the conditions specified in \cite[Lemma 2]{ACC_2025_Universal_Observer}. 
For the GPS-aided INS case, we consider the same measurement configuration as described in Section~\ref{subsection:convergence_analysis}.

Consider a vehicle moving in 3D space and tracking the following eight-shaped trajectory: 
$$p(t)=\begin{bmatrix}\cos(5t)\\ \sin(10t)/4\\ -\sqrt{3}\sin(10t)/4]\end{bmatrix}.$$ 
The rotational motion of the vehicle is subject to the following angular velocity:
$$ \omega(t)=\begin{bmatrix}\sin(0.3t) \\0.7\sin(0.2t+\pi) \\0.5\sin(0.1t+\pi/3)\end{bmatrix}.$$ The initial values of the true pose are $\posif(0)=\begin{bmatrix}
1 & 0 &0
\end{bmatrix}^{\top}$, $\velif(0)=\begin{bmatrix}
    -0.0125 & 2.5 & -4.33
\end{bmatrix}^{\top}$ and $R(0)=\exp([\pi e_2]_{\times}/2)$. The norm of the gravity vector $\gravifr$ is fixed at $9.81m/s^2$. The landmarks are located at $p^{\mathcal{I}}_{\ell_1}=[2\ 0\ 0]^{\top}$, $p^{\mathcal{I}}_{\ell_2}=[0\ 0.4\ 0]^{\top}$, $p^{\mathcal{I}}_{\ell_3}=[0\ 0\ 0.5]^{\top}$, $p^{\mathcal{I}}_{\ell_4}=[1 \ 0 \ 0 ]^{\top}$, $p^{\mathcal{I}}_{\ell_5}=[0\ 1 \  0]^{\top}$. 
The initial conditions for the observers are $\hat{p}^{\mathcal{I}}=\hat{v}^{\mathcal{I}}=\begin{bmatrix}
    1& 1& 1
\end{bmatrix}$, $\hat{e}_i^{\mathcal{B}}=e_i^{\mathcal{I}}$, $i\in\{1,2,3\}$, $\hat{R}(0)=\id$,  $P(0)=1\id$, $V(t)=10\id$,  $Q(t)=100\id$, $\rho_1=10$, $\rho_2=6$, $\rho_3=4$ . 
The magnetic field $\xi_1^{\mathcal{I}}$ is set to $[\begin{matrix} \tfrac{1}{\sqrt{2}} & 0 & \tfrac{1}{\sqrt{2}} ]\end{matrix}^{\top}$. The measurements are considered to be affected by a Gaussian noise of a noise-power as follows : $10^{-1}$
for IMU and magnetometer measurements  and  $5.10^{-2}$ for Stereo measurements. The simulation results are presented in Figs.~\ref{figure:simulation_result_stereo} and~\ref{figure:simulation_result_GPS}. It can be clearly seen that the estimated trajectories converge to the true trajectory after some seconds. Overall, the observer demonstrates a good noise-filtering capability.

\begin{figure}[h!]
    \centering
\includegraphics[width=\columnwidth]{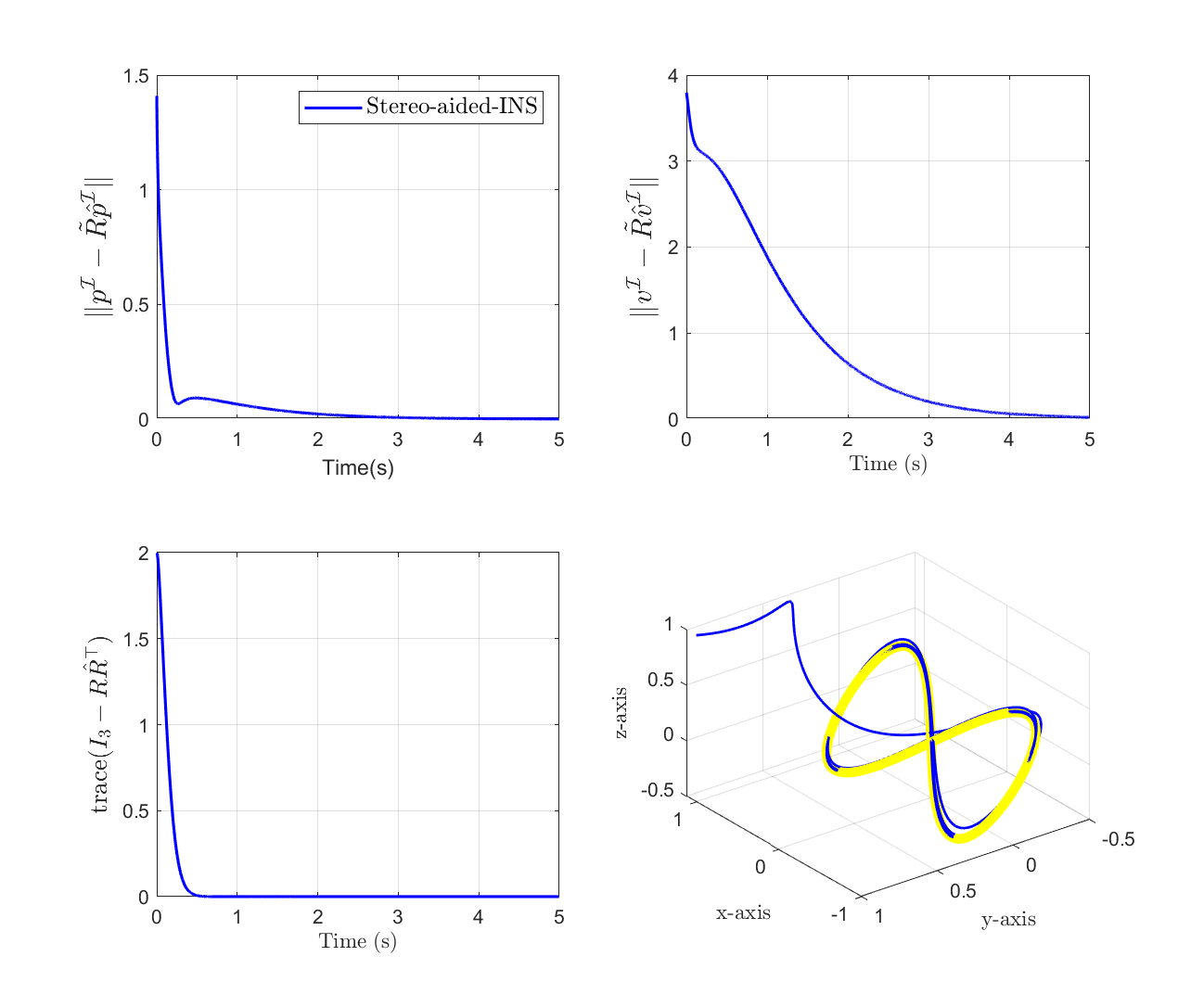}
    \caption{Estimation errors and trajectories for Stereo-aided INS.}
    \label{figure:simulation_result_stereo}
\end{figure}


\begin{figure}[h!]
    \centering
\includegraphics[scale=0.4]{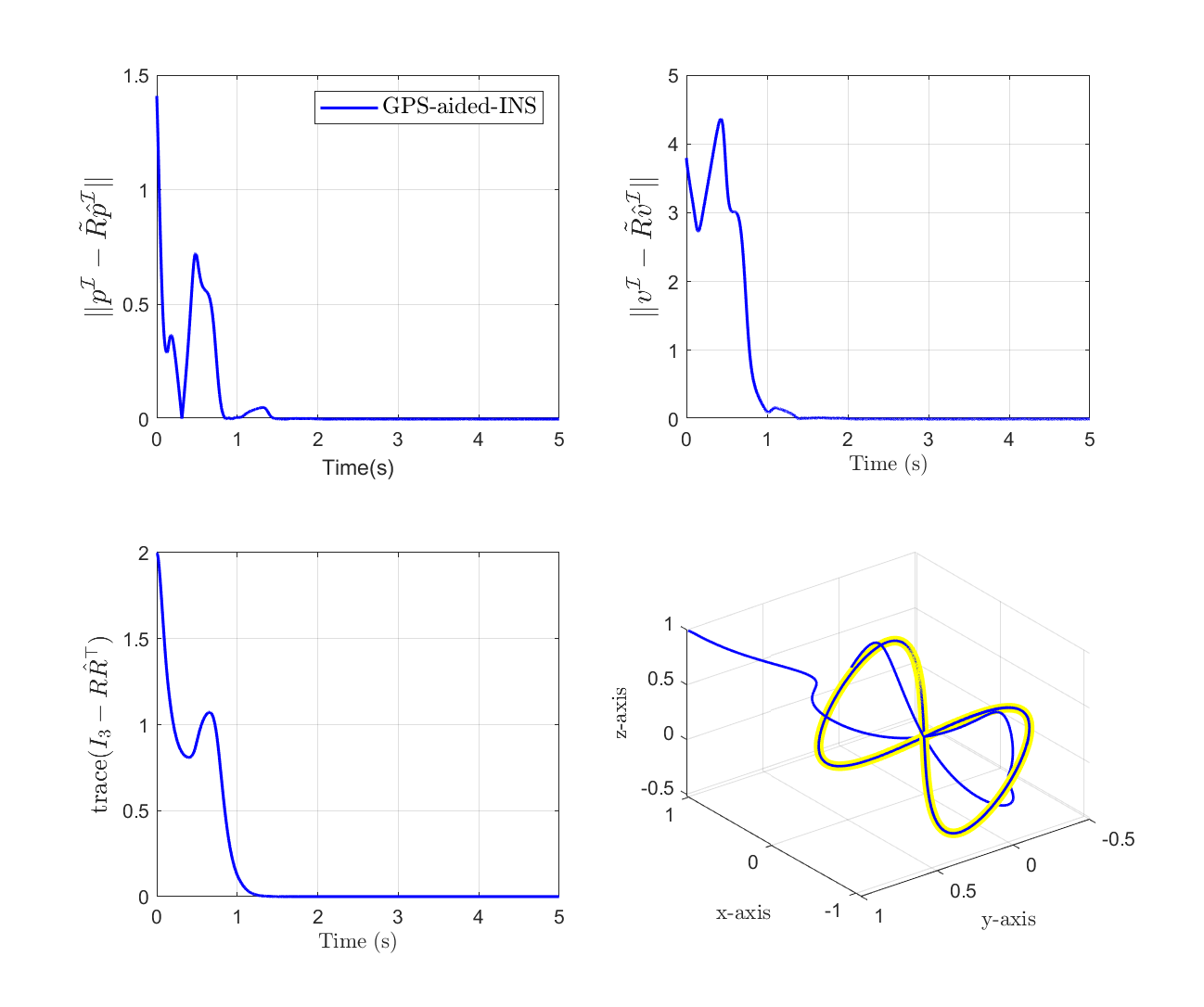}
    \caption{Estimation errors and trajectories for GPS-aided INS.}
    \label{figure:simulation_result_GPS}
\end{figure}

\section{Conclusion}\label{section:conclusion}

In conclusion, this paper introduces a novel nonlinear geometric observer on $\sefive$ for inertial navigation, achieving almost global asymptotic stability (AGAS). By embedding the system in an extended state space with three auxiliary variables, we are able to formulate a decoupled error dynamics structure that supports a wide range of inertial-frame and body-frame measurements. This embedding enables the reformulation of inertial-frame measurements as body-frame relative measurements, fitting seamlessly within the $\sefive$ observer design framework and allowing for simplified gain design, as the translational error dynamics align with the trajectory of a continuous-time Kalman filter.

We recognize that this observer operates on a state space with a higher dimensionality than that of the original system, which imposes somewhat stronger uniform observability conditions for convergence. However, as discussed in \cite{ACC_2025_Universal_Observer}, by leveraging the orthogonality constraint \( RR^\top = I \) of the rotation matrix, it is possible to introduce auxiliary (pseudo) measurements that effectively relax these observability conditions, making them consistent with the minimum requirements necessary for robust state estimation. Future work will focus on extending this work to accommodate biased sensor measurements.
\section*{Appendix}

\subsection{Proof of Proposition~\ref{proposition:reformuulation_of_the_measurements}}\label{appendix:proof_of_proposition:reformuulation_of_the_measurements}
First, we recall that 
\begin{flalign*}
X^{-1}=\left[\begin{array}{c:c}
R^{\top}  & 
   \begin{matrix}
      -[R^{\top}\posif\; R^{\top}\velif \;R^{\top}e_1  \;R^{\top}e_2\;R^{\top}e_3]
   \end{matrix} \\
  \hdashline \\  [-1em] 
  0_{5\times 3}  & I_{5\times 5} \\
\end{array}
\right]
\end{flalign*}
We now consider four cases depending on the type of the considered output.\\
\textbf{Case 1:} Let $r_i^{\top}=\begin{bmatrix}          \gamma_i &0 &-(\xi_i^{\mathcal{I}})^{\top}\end{bmatrix}^{\top}$, we have for any $i\in\{1, \cdots,p\}$,
\begin{flalign}
X^{-1}\mathbf{r}_i=X^{-1}\begin{bmatrix}
       0_{3\times1}\\\gamma_i\\0\\-\xi^{\mathcal{I}}
\end{bmatrix}=\begin{bmatrix}  \eta^{\mathcal{B}}_i\\\gamma_i\\0\\-\xi^{\mathcal{I}}
\end{bmatrix}
\nonumber,
\end{flalign}
from which we obtain,
\begin{flalign}
X^{-1}\mathbf{r}_i=\begin{bmatrix}  y_i\\r_i
\end{bmatrix}=\mathbf{y}_i.
\end{flalign}
\textbf{Case 2:} Similarly, define $r_i^{\top}=\begin{bmatrix}
           1 &0&-(\boldsymbol{\eta}_i^{\mathcal{I}})^{\top}
      \end{bmatrix}^{\top}$, we have for any $i\in\{p+1, \cdots,p+q\}$,
\begin{flalign}
X^{-1}\mathbf{r}_i=X^{-1}\begin{bmatrix}  0_{3\times1}\\1 \\0\\-\boldsymbol{\eta}_i^{\mathcal{I}}
\end{bmatrix}=\begin{bmatrix}   b_{i-p}\\1 \\0\\-\boldsymbol{\eta}_i^{\mathcal{I}}
\end{bmatrix},
\end{flalign}
then, we have \begin{flalign}
X^{-1}\mathbf{r}_i=\begin{bmatrix}  y_i\\r_i
\end{bmatrix}=\mathbf{y}_i.
\end{flalign}
\textbf{Case 3:} Let $r_i^{\top}=\begin{bmatrix}
            0 &1&-(\eta_v^{\mathcal{I}})^{\top}
      \end{bmatrix}^{\top}$, we have for  $i=p+q+1$,
\begin{flalign}
X^{-1}\mathbf{r}_i=X^{-1}\begin{bmatrix}  0_{3\times1}\\  0 \\1\\-\eta_v^{\mathcal{I}}
\end{bmatrix}=\begin{bmatrix}  0_{3\times1}
\\  0 \\1\\-\eta_v^{\mathcal{I}}
\end{bmatrix},
\end{flalign}
which leads to\begin{flalign}
X^{-1}\mathbf{r}_i=\begin{bmatrix}  y_i\\r_i
\end{bmatrix}=\mathbf{y}_i.
\end{flalign}
\textbf{Case 4:} Take $r_i^{\top}=\begin{bmatrix}
            0 &-1&0_{1\times3}
      \end{bmatrix}^{\top}$, we have for  $i=p+q+2$,
\begin{flalign}
X^{-1}\mathbf{r}_i=X^{-1}\begin{bmatrix}  0_{3\times1}\\  0 \\-1\\0_{3\times 1}
\end{bmatrix}=\begin{bmatrix} \eta_v^{\mathcal{B}}\\  0 \\-1\\0_{3\times 1}
\end{bmatrix},
\end{flalign}
this yields\begin{flalign}
X^{-1}\mathbf{r}_i=\begin{bmatrix}  y_i\\r_i
\end{bmatrix}=\mathbf{y}_i. 
\end{flalign}
Therefore, \eqref{equation:universal_form_for_output_in_SEfive} is verified for any $i\in\{1,\cdots,m\}$, which concludes the proof.

\subsection{Proof of Lemma~\ref{lemma:dynamics_of_the_translaional_error_in_the_body_frame}}\label{appendix:proof_of_lemma_dynamics_of_the_translaional_error_in_the_body_frame}
  We begin by expressing the dynamics of $\tzB$ as follows:
   \begin{equation}
       \dot{\tilde{z}}^{\mathcal{B}}=-\wx\tzB+\tzB\bar{A}^{\top}-\hat{R}^{\top}K(I_5\otimes \Delta_z),
   \end{equation}
   which leads to the following expression for the dynamics of $\tilde{x}^{\mathcal{B}}$,
     \begin{equation}\label{equation:dynmaic_of_x_B}
\dot{\tilde{x}}^{\mathcal{B}}=A(t)\tilde{x}^{\mathcal{B}}+\vect(\hat{R}^{\top}K(I_5\otimes \Delta_z)),
   \end{equation}
   where $A(t)=\bar{A}\otimes I_3+S(t)$ and $S(t)=
-I_5\otimes\wx$. Moreover,  by applying the identity $\vect(ABC) = (C^{\top} \otimes A) \vect(B)$, it follows that:
\begin{flalign}\label{equation:vec(RK(IoDz)}
   \vect(\hat{R}^{\top}K(I_5\otimes \Delta_z))&=(I_5\otimes \hat{R}^{\top})\vect(K(I_5\otimes \Delta_z))\nonumber\\
   &=(I_5\otimes \hat{R}^{\top})\begin{bmatrix}
       K_p\\K_v\\K_{e1}\\K_{e2}\\K_{e3}
   \end{bmatrix}\Delta_z.\\
   &=(I_5\otimes \hat{R}^{\top})K^{\mathcal{I}}\Delta_z.\nonumber
\end{flalign}
On the other hand we have $G\Delta \mathbf{y}_i=\vect(G\Delta \mathbf{y}_i)=\vect(\tilde{R}^{\top}\tilde{z}r_i)$. Since $\tilde{R}^{\top}=\hat{R}R^{\top}$ and by applying the identities $\vect(ABC) = (C^{\top} \otimes A) \vect(B)$ and $(A\otimes B)(C \otimes D) = (AC)\otimes(BD)$, we obtain 
\begin{align}
 G\Delta \mathbf{y}_i&=(r_i^{\top}\otimes \hat{R})\vect(R^{\top}\tilde{z})\\
 &=\hat{R}(r_i^{\top}(t)\otimes I_3)\tilde{x}^{\mathcal{B}}, 
\end{align}
from which it follows that,

\begin{align}\label{equation:Dz_in_funcition_of_xB}\begin{split}
\Delta_z&=(I_m\otimes\hat{R})\begin{bmatrix}
 (r_1^{\top}(t)\otimes I_3)   \\
 \vdots \\
 (r_m^{\top}(t)\otimes I_3)
\end{bmatrix}\tilde{x}^{\mathcal{B}}\\
 &=(I_m\otimes\hat{R})C(t)\tilde{x}^{\mathcal{B}}.\end{split}
\end{align}
Hence, in view of \eqref{equation:Dz_in_funcition_of_xB},  \eqref{equation:vec(RK(IoDz)} yields,
\begin{flalign}\label{equation:K_BC(t)x_B}
   \vect(\hat{R}^{\top}K(I_5\otimes \Delta_z))&=(I_5\otimes \hat{R}^{\top})K^{\mathcal{I}}(I_m\otimes\hat{R})C(t)\tilde{x}^{\mathcal{B}},\nonumber\\   &=K^{\mathcal{B}}(t)C(t)\tilde{x}^{\mathcal{B}}.
\end{flalign}
Combining \eqref{equation:dynmaic_of_x_B} and \eqref{equation:K_BC(t)x_B}, we obtain \eqref{equation:closed_loop_dynamic_of_x_B}, which concludes the proof.

\subsection{Proof of Theorem~\ref{theorem_AGAS}}\label{appendix:proof_of_theorem_AGAS}
We first rewrite the innovation term $\Delta_R$ as follows:
\begin{equation}\label{equation:innovation_term_rewritten_as_two_terms}
\Delta_R=\psi(M\tilde{R})+\Gamma(\hat{R})\tilde{x}^{\mathcal{B}},
\end{equation}
where    \begin{flalign*}
\hspace{0.4cm} M&=\mathrm{diag}(\rho_1,\rho_2,\rho_3),\;
\psi(M\tilde{R})=-\frac{1}{2}\sum_{i=1}^3\rho_i[e_i]_{\times}\tilde{R}^{\top}e_i,
\\
 \Gamma(\hat{R})&=\frac{1}{2}\begin{bmatrix}
    0_{3\times6}, \rho_1[e_1]_\times\hat{R}^{\top} ,\rho_2[e_2]_\times\hat{R}^{\top}, \rho_3[e_3]_\times\hat{R}^{\top}
\end{bmatrix}.\nonumber
\end{flalign*}
Hence, the closed loop dynamics of the geometric error can be written in an explicit form as follows:
\begin{align}
    \dot{\tilde{R}}&=\tilde{R}[-\psi(M\tilde{R})-\Gamma(\hat{R})\tilde{x}^{\mathcal{B}}]_\times,\label{equation:explicit_closed_loop_system_attitude}\\
    \dot{\tilde{x}}^{\mathcal{B}}&=(A(t)+K^{\mathcal{B}}(t)C(t))\tilde{x}^{\mathcal{B}}.\label{equation:explicit_closed_loop_system_translation}
\end{align}
Note that the above closed-system can seen as a cascade interconnection of a linear time-varying (LTV) system on $\R^{15}$ \eqref{equation:explicit_closed_loop_system_attitude} and a nonlinear system evolving on $\so$ \eqref{equation:explicit_closed_loop_system_attitude}. Therefore, to establish the desired result, we begin by demonstrating that the attitude subsystem \eqref{equation:explicit_closed_loop_system_attitude} is almost ISS at $I_3$ with respect to $\tilde{x}^{\mathcal{B}}$. Since the pair $(A(t),C(t))$ is uniformly observable,  \eqref{equation:explicit_closed_loop_system_translation} is GES, and thus there exists a closed set $\mathcal{S}\subset\R^{15}$ such that $\tilde{x}^{\mathcal{B}}(t)\in \mathcal{S}$, for any $t\geq0$. Additionally, we have  $\Gamma(\hat{R})$ is bounded since $\textstyle \|\Gamma(\hat{R})\|_F\leq\frac{\sqrt{2}}{2}\sum_{i=1}^{3}\rho_i$, for any $\hat{R}\in\so$, and the matrix $M$ is positive definite with three distinct eigenvalues, $\rho_1$, $\rho_2$ and $\rho_3$. Hence, conditions of \cite[Proposition 1]{Wang_2024_TAC_Intermittent_Measurements} are satisfied by taking $\bar{\Gamma}(\hat{R})=-\Gamma(\hat{R})$, $\textstyle c_{\bar{\Gamma}}=\frac{\sqrt{2}}{2}\sum_{i=1}^3\rho_i$, $A=M$, $k_o=1$,  and thus the attitude subsystem \eqref{equation:explicit_closed_loop_system_attitude} is ISS at $I_3$ with respect to $\tilde{x}^{\mathcal{B}}$. Therefore, given that \eqref{equation:explicit_closed_loop_system_attitude} is ISS and \eqref{equation:explicit_closed_loop_system_translation} is GES, it follows from  \cite[Theorem 2]{Angeli_ISS_plus_GES_gives_AGS} that the equilibrium point $(I_3,0_{15\times 1})$ is AGAS, which concludes the proof.

\bibliographystyle{unsrt}
\bibliography{main}
\end{document}